\documentclass[]{elsart}
\usepackage{graphicx, amssymb}
\begin{document}

\begin{frontmatter}
\title{Efficient Parallel Algorithm for Statistical Ion Track Simulations
in Crystalline Materials}
\author{Byoungseon Jeon \corauthref{cor}}
\author{Niels Gr{\o}nbech-Jensen}
\corauth[cor]{Corresponding author}
\address{Department of Applied Science, University of California, 
Davis, California 95616}

\date{\today}

\begin{abstract} 
We present an efficient parallel algorithm for statistical Molecular
Dynamics simulations of ion tracks in solids. The method is based on
the Rare Event Enhanced Domain following Molecular Dynamics (REED-MD)
algorithm, which has been successfully applied to studies of, e.g.,
ion implantation into crystalline semiconductor wafers.
We discuss the strategies for parallelizing the method, and we settle
on a host-client type polling scheme in which a multiple of asynchronous
processors are continuously fed to the host, which, in turn, distributes
the resulting feed-back information to the clients. This real-time
feed-back consists of, e.g., cumulative damage information or statistics
updates necessary for the cloning in the rare event algorithm.
We finally demonstrate the algorithm for radiation effects in
a nuclear oxide fuel, and we show the balanced
parallel approach with high parallel efficiency in multiple 
processor configurations.
\end{abstract}


\end{frontmatter}

\newenvironment{sourcesample}
{\begin{list}{}{\setlength{\leftmargin}{1em}}\item\scriptsize\bfseries}
{\end{list}}

\section{Introduction}
 Numerical simulations of high energy ion tracks and range distributions
in solids are limited by several factors, including the total range
of the ion and the small time step necessary for resolving high energy 
atomic collisions.  These problems have been successfully addressed for 
disordered or dense materials through event-based Binary Collision (BC) 
algorithms that simulate straight atomic trajectories between instantaneous 
pairwise collisions with the target material \cite{Robinson_74}.
 As an integral part of this approach, the BC model relies on simulating 
only a very small sample of the target material at any given time, since 
the interaction between a moving ion and the material must be short range 
in binary collisions.  The combination of these two features allows for 
studies of very high energy ionic paths over long range, provided that
the interactions are well represented as binary collisions and that simple
relationships between the trajectory can be established easily and accurately,
before and after a collision.  A Molecular Dynamics (MD) revision to the 
binary collision technique  has been introduced in order to capture
the propagation of an ion through crystalline materials in which the crystal
structure may provide for the channeling of ions or other structural effects
that make the binary collision approximation inappropriate \cite{beardmore98}.
This approach requires more computational resource than BC 
\cite{marlowe, srim1, srim2} in advancing the traveling ion while all the 
relevant atomic interactions are included every time step. 

However, this MD approach provides more correct representation of channel 
symmetry which a strict binary collision may not describe, and also the 
computational efficiency can be spared by the help of parallel computing.
The strategy is therefore to model all the relevant interactions between
a moving ion and the surrounding material temporally continuous, while
simultaneously making the essential assumption that an ion propagates 
through the target material only guided by its immediate spatial vicinity;
i.e., a domain following approach that we retain from BC.  Thus, this 
methodology is a hybrid between BC and MD approaches.  Consequently, the 
enhanced accuracy of the hybrid is at the price of computational  speed, 
requiring a larger number of calculations for the simulations of collisions 
or  interactions. Further combining the domain following MD scheme with 
a rare-event enhancing algorithm, which can efficiently evaluate ion 
probabilities (e.g., as a function of range) over several orders of magnitude,
produced the Rare Event Enhanced Domain following Molecular Dynamics (REED-MD)
scheme \cite{beardmore98}.  This method has successfully produced accurate 
dopant density profiles of ion-implanted semi-conductor wafers 
\cite{cai98,beardmore99} at ion energies in the range of 10keV-100keV 
initial energy with minimal empirical fitting of model parameters.  The 
approach has subsequently been reproduced by other groups 
\cite{sillanpaa99,kang00} who also studied semiconductor doping by 
ion irradiation.

The recent resurgence of interest in nuclear materials provides a new class 
of applications of the REED-MD approach to ion track simulations. For example,
a typical fuel is actinide-oxide (e.g., UO$_{2+x}$ or 
U$_y$Pu$_{1-y}$O$_{2\pm x}$ or U$_y$Pu$_z$Am$_{1-y-z}$O$_{2\pm x}$)
\cite{Olander,VanBrutzel06} in which either spontaneous decay produces 
high energy ion recoils of $\sim$85keV or neutron induced fission produces 
ion energies of order 85MeV.  Understanding the evolution of both the ion 
paths and the long time properties of the target material demands practical 
simulation tools that can simulate a broad spectrum of very high energy ions 
in crystalline structures with a variety of damage, imperfections, and 
structures \cite{Olander}. The very high energies demand both fine time step 
resolution and, for light elements, simulations of relatively long 
trajectories (up to several $\rm \mu m$).  In order to mitigate this 
computational load we propose to take advantage of parallel computing of 
individual ion tracks.  Since both computational and physical behavior
of the different ion tracks are connected through possible damage
cumulation as well as the rare event statistical enhancement, the
parallel strategy is non-trivial. The aim of this paper is therefore to
demonstrate a parallel simulation method for efficiently evaluating the
statistics of ion ranges, and we exemplify the method through simulations
of uranium recoil and fission fragment propagation in crystalline uranium
di-oxide.

\section{Review of the REED-MD Algorithm}
In order to describe the parallelization of the REED-MD method, we
first briefly outline the core of the approach.

\begin{figure}
\centering
\includegraphics[clip,scale=0.25, angle=0]{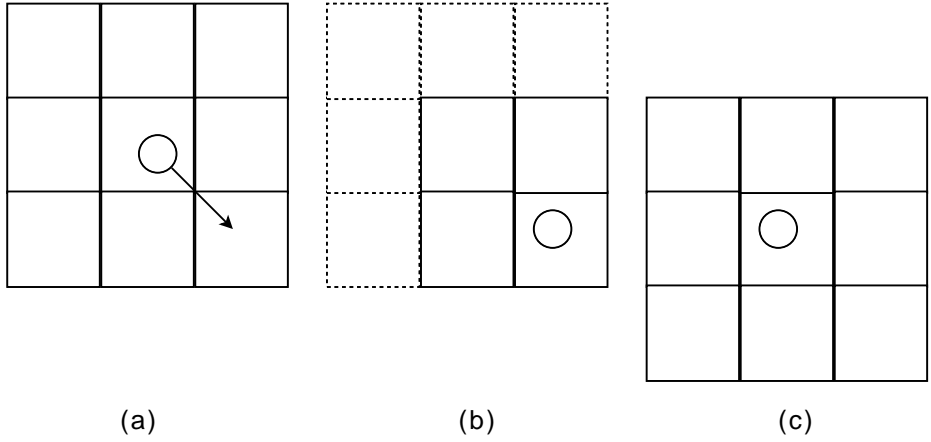}
\caption{If the ion breaches neighboring cells, then main cell moves to the
corresponding cell. (a) check the ion breach (b) moving the main cell and
remove next-neighboring cells (c) produce new neighboring cells.}
\label{fig:moving_cell}
\end{figure}

\begin{figure}
\centering
\includegraphics[clip, scale=0.5, angle=0]{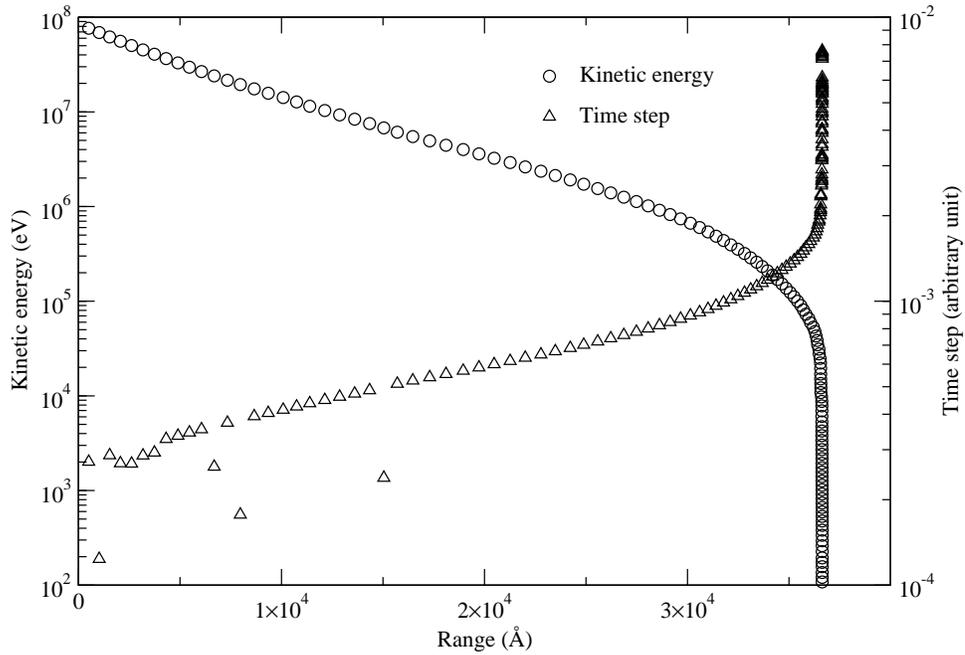}
\caption{A sample result of the radiation range test by Kr 85Mev ion with
UO$_2$ lattices. As kinetic energy decreases, an appropriate time step
increases by several orders of magnitude when using the adaptive time step
control \cite{beardmore98}. The exact unit of the time step is 10.18 fs.}
\label{fig:dt_E}
\end{figure}

Based on a purely classical MD formalism, REED takes advantage of several
particularities of high-energy ion collisional transport in order to
improve efficiency \cite{beardmore98}. {\bf First}, at kinetic or collisional
energies significantly larger than chemical or ionic bond strength, we do not
consider electronic structure details of the atomic interactions.
Instead, only the short range universal Ziegler-Biersack-Littmark (ZBL)
\cite{zbl} screening of nuclear charges is included. {\bf Second}, since 
the ZBL screening function provides for only short range interactions, 
a moving ion only experiences interactions with its immediate surroundings, 
which then allows for simulating only a small shell of atoms near the moving 
ion. As the ion propagates through the target material the simulated atoms 
are discarded in its wake and new material is created in front of the ion 
such that the added material correctly represents the desired statistical 
structure of target material configurations. The amount of simulated sample 
material depends on the interaction ranges between the moving ion and the 
target material. For UO$_2$ target material, we have settled on 27 unit 
cells of the 12-atom fluoride structure with the ion occupying the center 
unit cell. This domain following scheme is illustrated in 
Fig.~\ref{fig:moving_cell}. {\bf Third}, deformations of the electronic 
structure during collisions, as well as in interstitial regions, are modeled 
through the local inelastic interaction given by Firsov and Kichenevskii 
\cite{elteckov} and the Brandt-Kitagawa like electronic stopping mechanism 
\cite{BK,cai96} from the interstitial electron density, which is calculated 
by the residual unbound electrons in a muffin-tin model of the target 
material. Following the outline above, we are able to accurately
simulate the path of a fast-moving ion in a material for which we statistically
know the structure. These components are incorporated into a Newtonian 
equation of motion, which is simulated by a standard Verlet-type numerical 
integrator \cite{Verlet} with adaptive time-step control. We note that the 
dynamics is non-relativistic for the relevant energies and ions, and both 
the electronic stopping and the ZBL screening function are within the 
assumed validity ranges \cite{zbl}. Yet, because of the high initial kinetic 
energy, a very small time step is required to describe the initial behavior. 
But as the ion is decelerated, the appropriate time step increases by
several orders of magnitude, as shown in Figure \ref{fig:dt_E}. 
The adaptive time step control is determined based on both kinetic energy
and collisional potential energy. Detailed expression can be found in
Ref.~\cite{beardmore98}.

In order to obtain, e.g., a statistical range density (probability)
profile of a given type of ion in a material under given circumstances,
we must simulate a significant number of ion tracks, which differ in
the statistical realization of temperature and material defects.
A typical profile of this kind is characterized by a relatively short
range density peak (nuclear collisions) with less likely tails of deeper 
penetration (channeling). Thus, if one wishes to generate statistics with 
a given small variance in the deep range of ion tracks, a very large number 
of simulations must be produced, since most of these will contribute only 
to the primary range peak.

Alleviating this statistical problem, a Rare Event statistical Enhancement
algorithm \cite{beardmore98} is applied to provide uniform simulation effort at
all relevant ranges in order to equalize the variance of the density
distribution. This is accomplished by cloning the simulation system if
an ion reaches certain ranges. These ranges are determined such that
the acquired statistical count in each range interval between cloning
distances is equal (or similar). The correct statistical measure is then
recovered by consistently assigning a statistical weight to each simulated
ion; i.e., the statistical weight of a simulated ion is cut in half
every time it has been cloned. 
A simulation is initiated with a certain standard guess of the distribution 
of cloning depths, or it can be obtained from some other source,
and these depths are subsequently dynamically adjusted throughout the
acquisition of statistics in order to reach the desired goal of
uniform statistical count in all intervals. We note that the trajectories of
two clones of the same ion will rapidly deviate due to the statistical
representation of, e.g., temperature and crystal defect content in the
target material that is created as each ion is independently progressing
after cloning. Following this procedure, REED-MD can accurately and
efficiently produce range profiles of energetic ions in crystalline materials
with uniform statistical uncertainty over many orders of magnitude.

Even with the abovementioned strategy there can be significant computational
shortcomings when simulating nuclear fuel materials or nuclear radiation. 
For our test example (UO$_2$) in this paper, the ion energies are very large, 
requiring very small numerical time steps and yielding relatively large ranges
($\sim\rm \mu m$ for 85MeV fission fragments, such as Kr and Ba). We may
therefore benefit from a parallel strategy of a REED-MD implementation.
The strategy we propose does not involve any additional assumptions or
limitations of the simulated physics, and the acquired statistics is the 
same as if one uses the serial REED-MD algorithm.

\section{Parallelization of REED-MD}
Because REED-MD relies on statistical sampling of possible ion trajectories
we initiate a large number of ions with the same statistical representation.
This provides good opportunities for parallel computing, which will now be
discussed.

Straightforward distribution of the initiated ions across available
processors seems an obvious strategy, but this simple approach comes with
a couple of complications unless special care is taken. The root of the
problem is the necessary accumulation of the statistics as the simulation 
progresses in order to appropriately account for, e.g., cumulative damage 
and in order to specify the proper locations of the cloning distances that 
result in optimal statistical accuracy. However, since each initiated ion 
results in an unpredictable path, which may represent a very broad distribution
of range and behavior, the cpu time needed to complete a simulation of a single
ion path can vary greatly. This is further amplified by the possibility
of $2^N$ clones from the $N$ cloning ranges in REED. Thus, collective 
(synchronous) message passing is not efficient, since this would lead 
to a broad range of idle times in all but the processor with most time 
consuming task.

One solution to the problem of the synchronization described is to
let all processors share a common file in lieu of direct communication.
This would serve as an asynchronous communication tool. In order to prevent
the possibility of simultaneous writing to the file by two different
processors, one would need to specifically address this issue together with 
the inherent inefficiency of disk access.

Another solution is Polling \cite{mpi}. We let each processor run its own
simulations without collective communication. Using {\bf PROBE}, any 
incoming data can be detected, and each processor can respond accordingly. 
An example of polling is shown in Table \ref{table:polling_basic}.
Even though implementation may not be straightforward, it is possible to
minimize the idling time of each processor during the extremely asynchronous
REED-MD simulations. We here employ a polling method extensively and
describe how it is designed and implemented through a host-client algorithm.

As discussed above, the success of a parallelization strategy lies in how
well we minimize the idling time of processors. The flow of the algorithm
should therefore be such that the processors are kept occupied by continuously
producing results from individual REED-MD tracks. Each ion is initiated with 
the current-time global statistics that determines the cloning criteria, 
which are updated gradually throughout the simulation based on all completed 
(i.e., stopped) ions. This adaptive refinement is continued until the end of 
the entire calculation. To accomplish this gradual update of the statistics, 
three types of communication are required: first, the results of individual 
ion simulations, such as ion ranges and their corresponding statistical 
clone-weights. Second, updated cloning criteria. Finally, the termination 
signal indicating that no new ions need to be initiated by a processor.

Through several tests, we found that a host-client algorithm is very 
effective for our purpose. We assign a single processor as a host   processor, 
and this processor handles most of the coordinating work, such as I/O,
update of cloning criteria, and management of simulation results.
The rest of the processors are devoted to ion simulations with minimized
load for communication. Polling is employed when the host and client processors
communicate, and the conventional message passing interface \cite{mpi} has 
been implemented, providing extensive portability across system architectures. 

When we implement a non-blocking {\bf PROBE}, we consider the required
frequency and priority, such that any completed ion results from a processor
can be received at any time. The non-blocking {\bf PROBE} is therefore
continuously active in the host processor while
the other processors send the data at the end of every single ion simulation 
including cloned ions. After the host processor receives enough simulation
results, it updates the cloning criteria. The host  must then send 
the updated criteria to all clients, and here polling is employed
again. To reduce the idling time of the host   processor, the clients employ
non-blocking {\bf PROBE} at every time step. It is important to note that
cloning criteria in the client nodes are only updated at the completion of
every ion track (including the cloned ions) in order to avoid corrupted or 
inconsistent cloning  by conflicting criteria during a track simulation.

\begin{figure}
\centering
\includegraphics[clip, scale=0.2, angle=0]{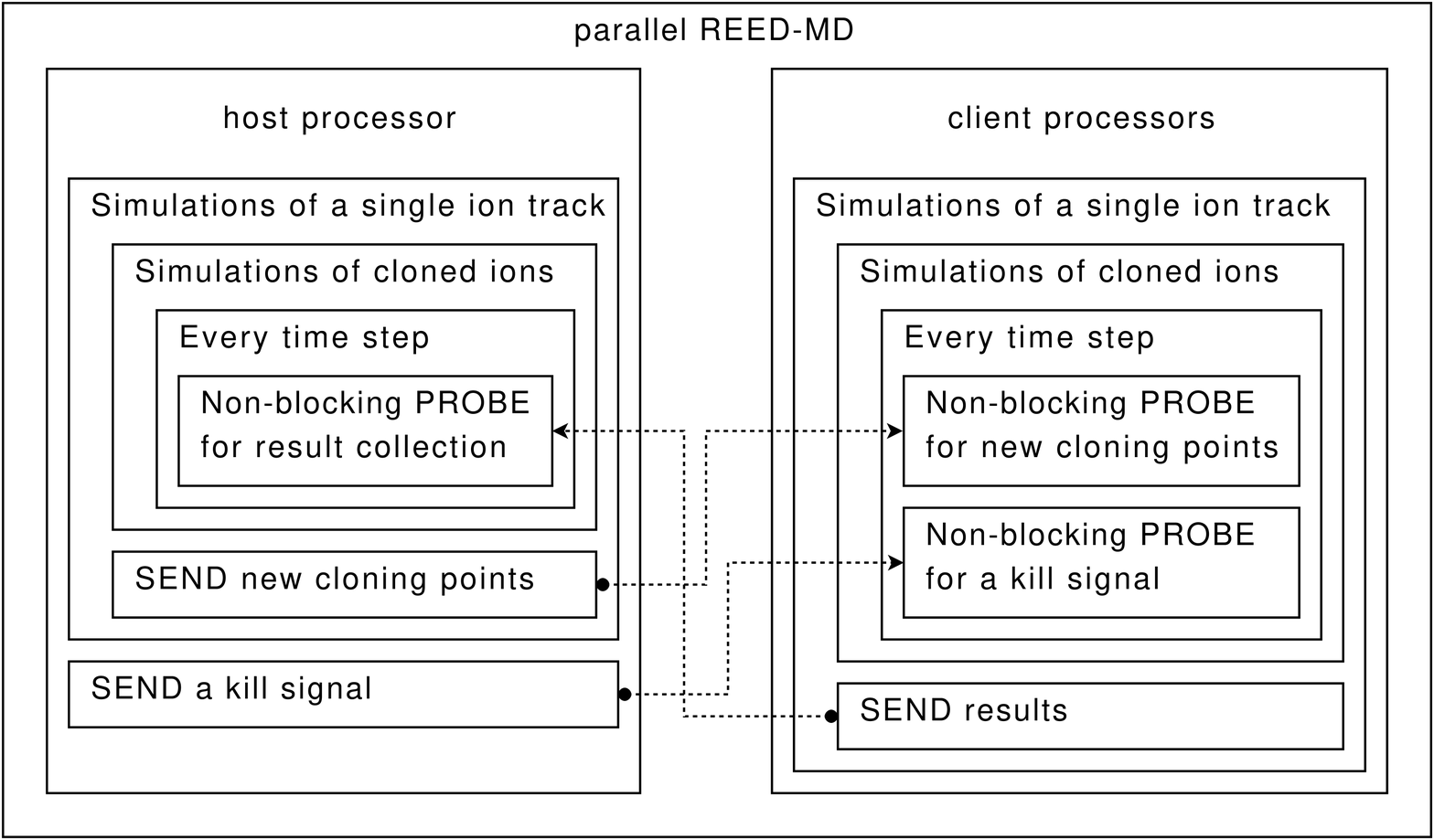}
\caption{Primitive host-client polling algorithm for REED-MD simulations. 
The locations of non-blocking {\bf PROBE} and {\bf SEND} routines are shown. 
For the host processor, sending new cloning
points might be performed per several ions, not every ion, in order to 
reduce communication load. }
\label{fig:polling}
\end{figure}

Finally, the host   processor will count the number of collected ions (or other
simulation statistics) to decide if the simulation should continue.
When a stop criterion is met at the host processor it
sends a kill-signal, and the client processors detect it by 
another non-blocking {\bf PROBE}, which is implemented to run every 
time step. This process is illustrated in Figure \ref{fig:polling}.

\begin{figure}
\centering
\includegraphics[clip, scale=0.2, angle=0]{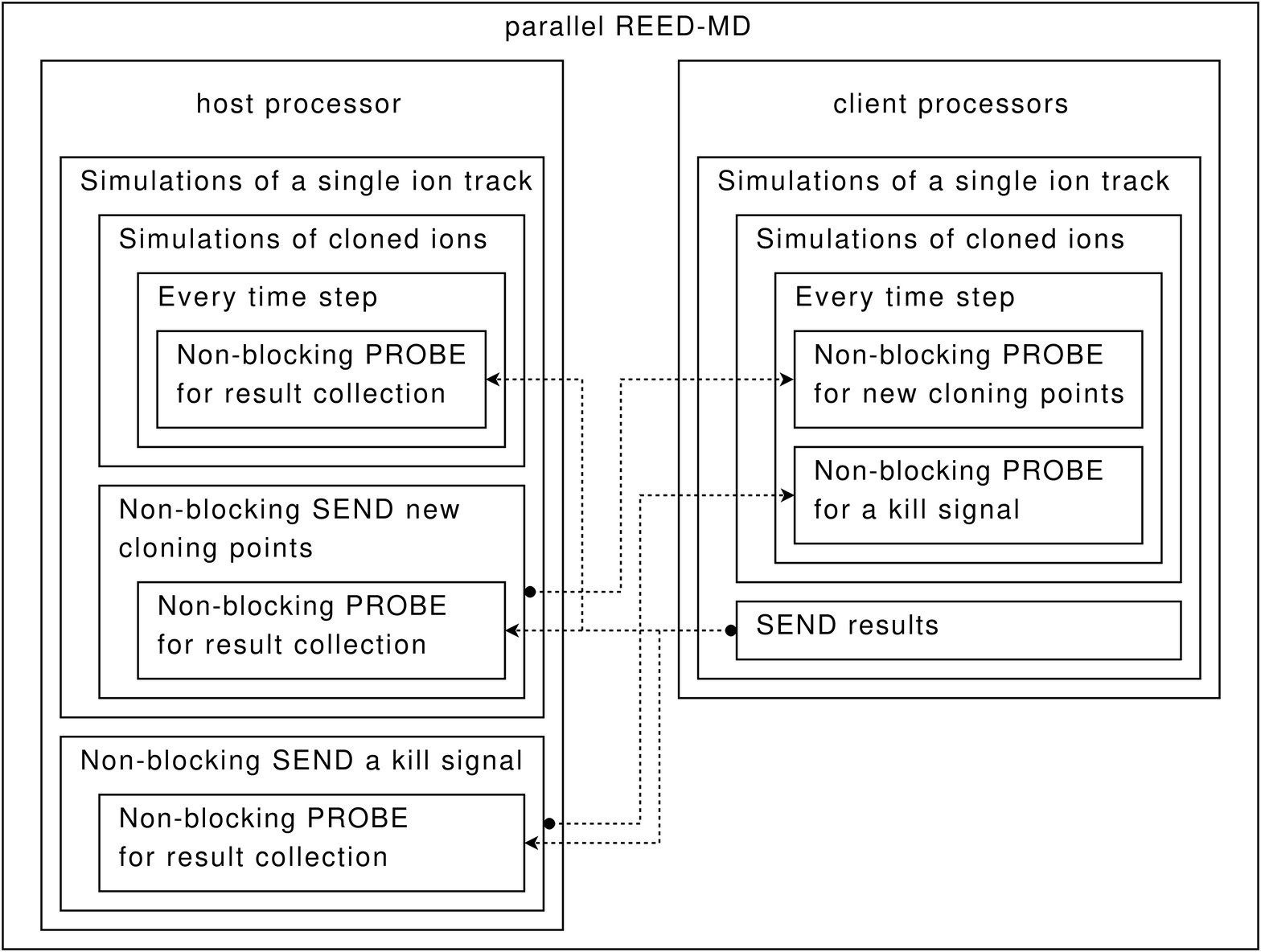}
\caption{Modified host-client polling algorithm. Non-blocking probe is
repeated in the send routines of the host   processor.}
\label{fig:polling2}
\end{figure}

Even though this method is straightforward and applies host-client
polling effectively, there is a serious problem. If both a host   and a client
processor send a signal simultaneously, waiting for the opposing
processor to respond, then the program may hang. Therefore, 
we need to modify the scheme using non-blocking {\bf SEND}.
As shown in Figure \ref{fig:polling2}, when the host   processor sends 
any data to the client processors, it employs non-blocking {\bf SEND}. 
Below the sending routine, we add one more non-blocking probe and a 
receive routine in order to prevent any immobilization  
of client processors. After receiving incoming data, non-blocking 
{\bf SEND} is completed. An example of the pseudo-code is shown in Table 
\ref{table:polling_adv}. 

\section{Tests of Parallel REED-MD}
We have built a parallel REED-MD code based on the developed algorithm, and
tested it on several configurations of multiple processors. We calculated 
the time consumption between routines, specifying computation and 
communication, and found the efficiency of the developed code. The tested 
machines are exemplified by a multiple of 2.5GHz quad-core G5 powerPC 
Macintosh nodes connected by a non-dedicated Gigabit network. The code was 
developed with the g++ 4.0.1 compiler and the Open MPI 1.1.5 library. 

Usually, parallel efficiency and scalability are measured relative to the
result of serial computing. However, due to the randomness of the 
simulation, computing costs may not be representative for a given
simulation, yielding the direct comparison to a serial simulation
somewhat ambiguous. Therefore, we focus on how much wall-clock time is 
devoted into the actual calculation of REED-MD, relative to the communication 
load and idling. This will serve as a measure of the efficiency of the 
developed parallel implementation.

\subsection{Radiation range by intermediate energy ions}
Our example target material is the fluoride structure of uranium dioxide
\cite{Olander}. We first show results of on-lattice uranium ions with initial
energy of 85keV and random direction, corresponding to the recoil energy
from a plutonium $\alpha$-decay. The target material has temperature
300K, simulated by a Gaussian distribution of the lattice atoms with RMS
values derived from full MD simulations of the equilibrium properties of
UO$_2$ \cite{Anurag08}. We complete simulations of approximately 5,000
initial ions, and we use eight dynamically optimized cloning distances.

\begin{figure}
\centering
\includegraphics[clip, scale=0.5, angle=0]{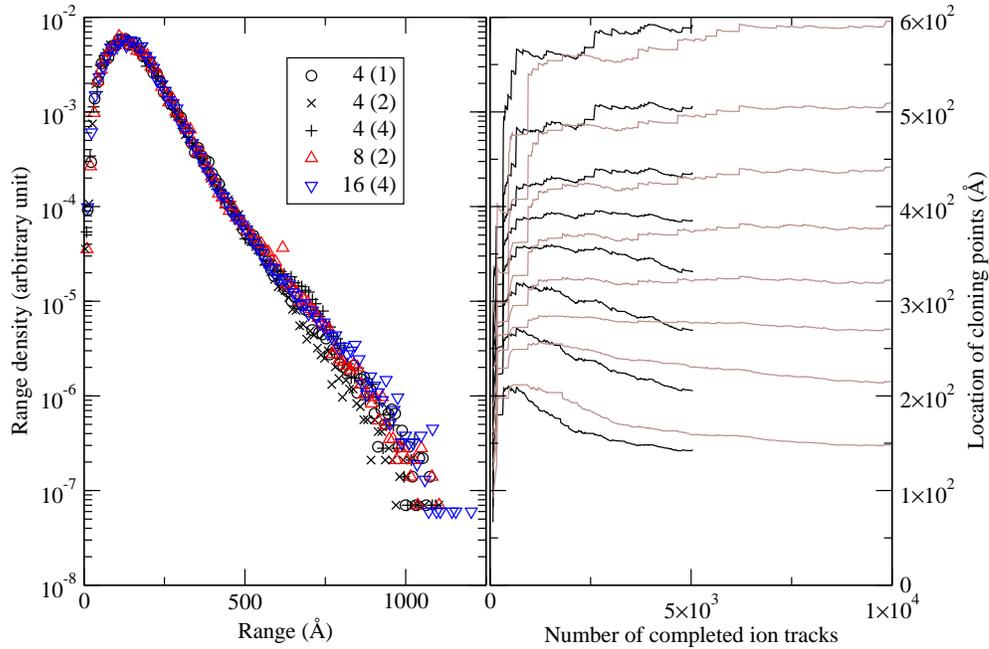}
\caption{Range results of U ion with 85 keV (left) and the evolution of
cloning points (right). Range density results are shown with the number
of the employed CPUs while the parenthesis shows the number of the employed
nodes. At first, 50 ion simulations were done without rare event enhancing, 
providing initial cloning criteria. Then 5,000 REED-MD simulations  
were  performed, updating the criteria gradually. Evolution curves (black)
are chosen from the simulations of 4 processor with 1 node as examples while
10,000 simulation results (gray) are provided as reference, showing saturated
evolution.}
\label{fig:test1}
\end{figure}

Range results and the evolution of cloning points are shown in Figure 
\ref{fig:test1}. We tested 1, 2, and 4 nodes corresponding to 4, 8, and 
16 CPUs, respectively. For all the parallel cases, range results of the left 
figure show good consistency, although slight discrepancies are found at the 
tail of the distribution. Such discrepancies must vanish for increasing 
number of initiated ions. The evolution of the cloning points are shown in 
the right figure and they are the results of 4 CPUs (single node) calculation.
We here see that the short range cloning distances stabilize sooner than the 
longer range ones, and we see that the longest ranges have not yet converged 
by the end of the simulation, indicating that the range distribution
has not yet acquired enough statistics at all scales. This is consistent
with the above observation of differences between the tails of the different
simulations using 5,000 initial ions. While the number of necessary initial
ions for reaching reliable statistics depends on both the physics of the 
system and the algorithmic number of cloning points (as well as the number
of orders of magnitude on which statistical data is desired), we generally
find that 10,000 initial REED-MD ions yield reliable statistics. The settling 
of all cloning points is a good indicator of adequate global statistics, and
we therefore use the cloning points as a criterion for determining the
quality of the simulation results.

The parallel performance of the procedure is summarized in 
Table \ref{table:parallel}. Here, computation load means the cost
devoted to the actual REED-MD calculations while communication load is
the cost of the communication with other processors. The results are
averaged over all the employed processors. Efficiency is the ratio of 
computation load to the total wall-clock time. The single node 4-CPU test, 
shows better than 99\% of wall-clock time devoted to actual REED-MD ion track
simulations. Basically, communication is very efficient within a single node, 
so optimal parallel performance may not be surprising in this case. However, 
the test demonstrates that potential parallel inefficiency of idling 
processors is not an issue in our implementation.
In order to study the effects of the networking, we repeated the 4-CPU
processor simulation with multiple nodes such as two and four. The results are
that the parallel efficiency drops from 99\% to 97\%, indicating the effect
of the gigabit connections and the latency of the networking switch. We then
conducted similar simulations with eight and sixteen CPUs (two and four nodes),
and the parallel efficiency remained 96\%. Since the amount of communicated
information is limited, we suspect that the drop in efficiency when the
simulations exceed one node is due more to the latency of the switch
than it is due to the speed of transmission. However, even 96\%
parallel efficiency is very useful for optimized computing.

The actual communication load of the host   processor was found to be
$2.5\times 10^2$, $2.1\times 10^3$, $2.0\times 10^3$, $3.3\times 10^3$, 
and $6.4\times 10^3$ seconds for 4(1), 4(2), 4(4), 8(2), and 16(4) CPUs 
(nodes), and the computing efficiency of the host   processor (measured as 
ion-track simulation time relative to total time) corresponds to 99.6\%,  
96.9\%, 97.1\%,  92.3\%, and 81.8\%. As the number of participating processors 
increases, the simulation results sent to the host   processor increase 
proportionally, and its local efficiency drops accordingly. However, with 
the efficiency of the client processors being nearly unchanged this strategy 
provides very high overall parallel efficiency at more than 96\%.

\subsection{Radiation range of high energy ions}
As an example of relevant fission energy simulations, we tested the code
performance for 85MeV Kr ions with up to 32 CPUs (8 nodes). 
The lattices and other boundary conditions are the same as in the above 
section. Compared to intermediate energy ions, the computational cost is 
very high, in part due to the small time step by the extremely high initial 
kinetic energy, and we tested only around a thousand initial ions.

\begin{figure}
\centering
\includegraphics[clip, scale=0.5, angle=0]{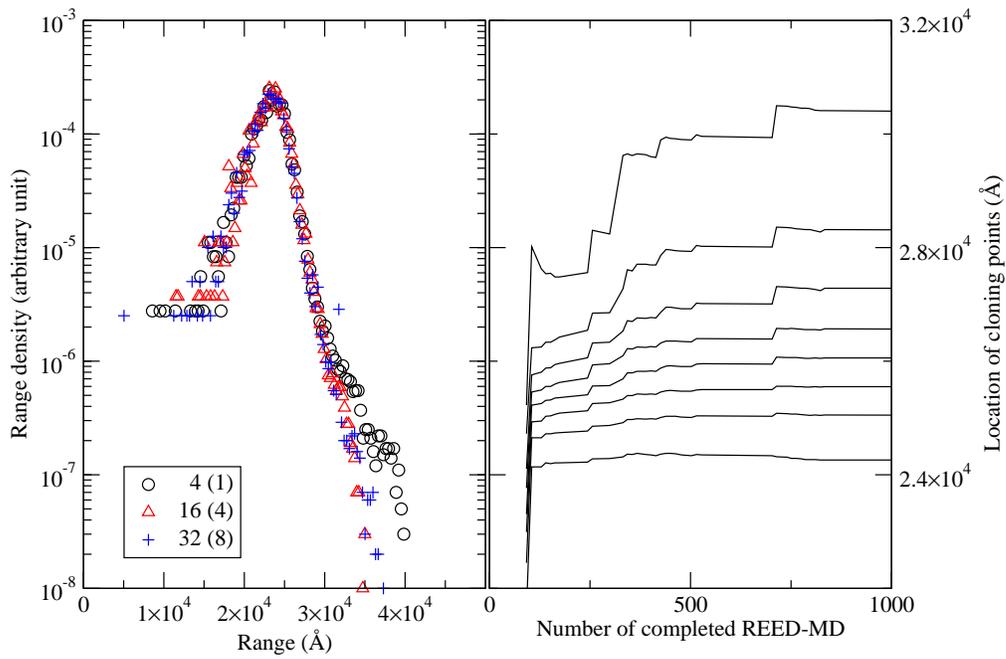}
\caption{Range results of Kr ion with 85 MeV (left) and the evolution of
cloning points (right). Notations are same as in Figure \ref{fig:test1}. 
Initial cloning criteria were determined from 50 ion simulations without rare event 
enhancing, then REED-MD were employed for the update of the criteria.}
\label{fig:test2}
\end{figure}

Range results and the evolution of the cloning points are shown in 
Figure \ref{fig:test2}. Because of the high energy, the ion track range
was found to reach up to several $\rm \mu m$. Each parallel run shows similar
range density but the tail of the distribution is quite different compared to 
above results. Because so few initial ions were simulated this discrepancy is
expected. As discussed above, more simulated ions will naturally result in
better statistics. Evolution curves of cloning points are from 4 (1) CPU
(node) run and they are still evolving, showing insufficient statistical 
sampling.

Parallel performance results are shown in Table \ref{table:parallel2}
and we find similar results to the abovementioned study of 85keV U,
providing better than 96\% efficiency. The communication load at the host   
processor was found as $9.3\times 10^2$, $1.7\times 10^4$, and 
$2.8\times 10^4$ seconds for 4(1), 16(4), and 32(8) CPUs(nodes), 
yielding 99.6\%, 82.2\%, and 62.6\% host node computing efficiency, 
respectively. As discussed above, more computing processors result in 
an avalanche of communication to and from the host   processor, and the 
host processor consequently suffers in its contributions to the data 
production; but overall performance still holds at better than 96\%. 
From these results, we can conclude that our host-client polling algorithm 
works well for parallel REED-MD simulations. Further, on dedicated servers 
with high speed networks, higher efficiency should be expected in line with 
our single node simulation results.

\section{Conclusion}
REED-MD is a robust method for simulating ionic radiation in dense and
crystalline materials. To refine the improbable statistics, rare event 
enhancement is employed and individual ion track simulations form different 
cascades and trajectories by cloning, resulting in asynchronous behavior 
among processors. Therefore, parallelization of the REED-MD
algorithm demands a strategy other than collective communication.

Based on the required priority and frequency of each communication,
we have designed and optimized appropriate algorithm locations 
for polling and message passing. Even with extremely asynchronous 
behavior among ion tracks of different processors, data could be
delivered efficiently between processors, leading to a nearly optimal parallel
efficiency.

We note that the success of providing this good parallel efficiency
even for severely asynchronous processor behavior implies that the 
algorithm is robust in its load balancing, which, in turn, implies that
the algorithm will perform equally well on a heterogeneous cluster of computing
nodes. We therefore envision that several other applications in Molecular 
Dynamics and Monte Carlo applications could take advantage of similar 
strategies.

\section*{Acknowledgments}
The authors are grateful for continuing useful discussions with Mark Asta and
Anurag Chaudhry.
This work was supported by the US Department of Energy Nuclear Energy
Research Initiative for Consortia (NERI-C) contract number
DR-FG07-071D14893 and in part by Los Alamos National Laboratory contracts
25110-001-05, through the Materials Design Institute, and
62410-001-08.


\clearpage

\begin{table}
\caption{Example of a polling routine for send/polling processors.}
\begin{center}
\begin{tabular}{|l|l|}
\hline
~~~polling processor & ~sending processor~ \\
\hline
\verb! MPI_Iprobe(..., flag, ...) ! & 
\verb! MPI_Send(...) !\\
\verb! if (flag) MPI_Recv(...)! & \\
\hline
\end{tabular}
\end{center}
\label{table:polling_basic}
\end{table}

\begin{table}
\caption{Example of a polling routine coupled with non-blocking sending.}
\begin{center}
\begin{tabular}{|l|l|}
\hline
~~~host    & ~client~ \\
\hline
\verb! MPI_Isend(..., request) ! & \verb! MPI_Recv(...) !\\
\verb!     MPI_Iprobe(..., flag, ...)! &  \verb! MPI_Send(...) ! \\
\verb!     if (flag) MPI_Recv(...) !   & \\ 
\verb! MPI_Waitall(..., request, ...)! & \\
\hline
\end{tabular}
\end{center}
\label{table:polling_adv}
\end{table}

\begin{table}
\caption{Results of parallel computing of REED-MD for a U ion with 85 keV. 
Number of tests is the number of simulated ions.
Results of computation and communication load are in units of seconds.}
\begin{center}
\begin{tabular}{r|r|r|r|r}
\hline
CPUs (nodes) & Number of tests & Computation  & Communication  & Efficiency\\
\hline\hline
 4 (1) & 5,053 & $6.7\times 10^4$ & $2.2\times 10^2$ & 99.7\% \\
 4 (2) & 5,057 & $6.7\times 10^4$ & $1.6\times 10^3$ & 97.7\% \\
 4 (4) & 5,072 & $6.8\times 10^4$ & $1.5\times 10^3$ & 97.8\% \\
 8 (2) & 5,094 & $4.5\times 10^4$ & $1.3\times 10^3$ & 97.1\% \\
16 (4) & 5,057 & $3.4\times 10^4$ & $1.2\times 10^3$ & 96.6\% \\
\hline
\end{tabular}
\end{center}
\label{table:parallel}
\end{table}

\begin{table}
\caption{Results of parallel computing of REED-MD for a Kr ion with 85 MeV. 
Notations are same as in Table \ref{table:parallel}.}
\begin{center}
\begin{tabular}{r|r|r|r|r}
\hline
CPUs (nodes) & Number of tests & Computation  & Communication & Efficiency\\
\hline\hline
 4 (1) & 1,142 & $2.5\times 10^5$ & $8.3\times 10^3$ & 99.7\% \\
16 (4) & 1,068 & $9.3\times 10^4$ & $3.1\times 10^3$ & 96.6\% \\
32 (8) & 1,222 &  7.3$\times 10^4$ & $2.6\times 10^3$ & 96.4\% \\
\hline
\end{tabular}
\end{center}
\label{table:parallel2}
\end{table}

\clearpage 

\begin{figure}
\centering
\includegraphics[clip, scale=0.25, angle=0]{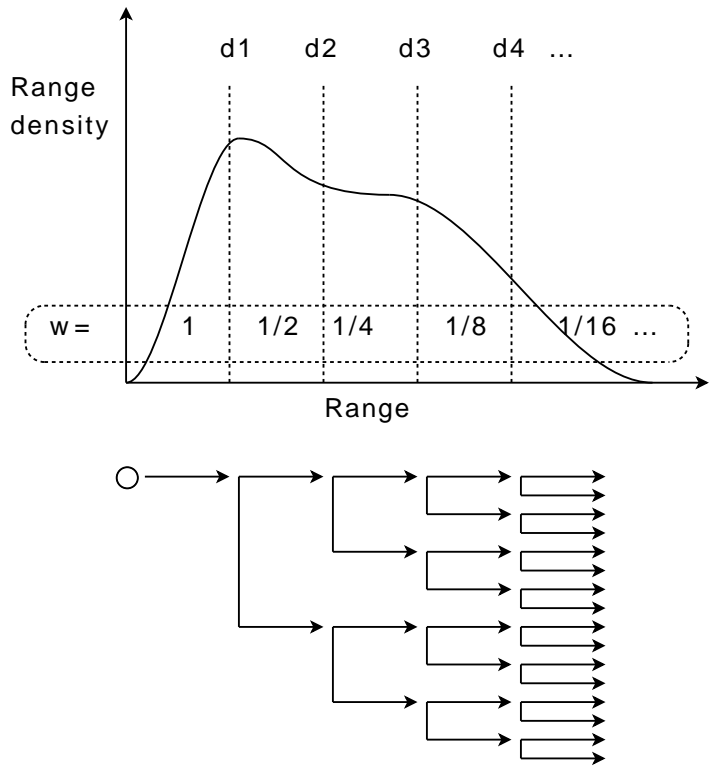}
\caption{Cloning of an ion simulation and corresponding weights.}
\label{fig:split}
\end{figure}

\end{document}